# Revealing the atomic structure of the buffer layer between SiC(0001) and epitaxial graphene

Sarah Goler [a,b], Camilla Coletti [a,c], Vincenzo Piazza [a], Pasqualantonio Pingue [b], Francesco Colangelo [b], Vittorio Pellegrini [b], Konstantin V. Emtsev [c], Stiven Forti [c], Ulrich Starke [c], Fabio Beltram [a,b], and Stefan Heun [b,*]

[a] *Center for Nanotechnology Innovation @ NEST, Istituto Italiano di Tecnologia, Piazza San Silvestro 12, 56127 Pisa, Italy*

[b] *NEST, Istituto Nanoscienze – CNR and Scuola Normale Superiore, Piazza San Silvestro 12, 56127 Pisa, Italy*

[c] *Max-Planck-Institut fuer Festkoerperforschung, Heisenbergstr. 1, 70569 Stuttgart, Germany*

**Abstract**

On the SiC(0001) surface (the silicon face of SiC), epitaxial graphene is obtained by sublimation of Si from the substrate. The graphene film is separated from the bulk by a carbon-rich interface layer (hereafter called the buffer layer) which in part covalently binds to the substrate. Its structural and electronic properties are currently under debate. In the present work we report scanning tunneling microscopy (STM) studies of the buffer layer and of quasi-free-standing monolayer graphene (QFMLG) that is obtained by decoupling the buffer layer from the SiC(0001) substrate by means of hydrogen intercalation. Atomic resolution STM images of the buffer layer reveal that, within the periodic structural corrugation of this interfacial layer, the arrangement of atoms is topologically identical to that of graphene. After hydrogen intercalation, we show that the resulting QFMLG is relieved from the periodic corrugation and presents no detectable defect sites.

---

[*] Corresponding author: Fax: +39-050-509-417.
E-mail address: stefan.heun@nano.cnr.it (S. Heun)

# 1. Introduction

The exploitation of graphene for practical applications crucially depends on methods for its controllable production on large areas. Presently, epitaxial growth of graphene on the Si-face of silicon carbide (*i.e.*, the SiC(0001) surface) is considered an extremely promising route for the production of large-area graphene suitable for device applications [1]. Indeed, growth of high-quality and homogeneous graphene layers was recently demonstrated [2,3]. Nevertheless, there is still conflicting theoretical and experimental evidence in the literature concerning the structural properties of the buffer layer *i.e.*, the carbon layer that resides below the monolayer of graphene on the SiC(0001) face. In scanning tunneling microscopy (STM) often a (6×6) corrugation is imaged [4-7]. However, this corrugation does not reflect the true $(6\sqrt{3}\times6\sqrt{3})R30^o$ periodicity that can be seen in low-energy electron diffraction (LEED) [8,9] and partially-resolved STM images [8]. The $(6\sqrt{3}\times6\sqrt{3})R30^o$ periodicity is imposed by the registry match of SiC and the carbon layer [10,11] and is predicted by theoretical calculations [12,13]. However, the structure of the buffer layer on an atomic scale is still debated since complete atomic resolution has eluded STM studies so far. Several reports suggested a hexagonal atomic arrangement such as in monolayer graphene [8,10-14], but also a nanomesh structure with isolated carbon islands was proposed [5]. The discovery that the buffer layer can be converted into pristine $sp^2$-bonded graphene (the so-called quasi-free-standing monolayer graphene (QFMLG)) via the intercalation of hydrogen at the interface with the SiC substrate (see sketch in the inset to Figure 2d) [15] was a strong argument in favour of the hexagonal atomic arrangement in the buffer layer. Nevertheless, recently an arrangement of hexagonal, pentagonal, and heptagonal atomic placements [16] was put forward as the buffer layer geometry. The clarification of the atomic structure of the buffer layer is imperative for understanding and controlling the epitaxial growth of mono- and few-layer graphene on SiC(0001). In this work we present atomically-resolved STM images of the buffer layer that clearly demonstrate a graphene-like honeycomb structure, hence resolving the dispute on the atomic arrangement within the interface layer. By exploiting scanning tunneling microscopy and spectroscopy we examine the atomic and electronic structure of the buffer layer and compare it to that of QFMLG.

## 2. Experimental

Buffer layer samples were grown by annealing atomically flat 6H-SiC(0001) samples in a radio-frequency (RF) induction furnace under an Ar atmosphere at about 1400$^{o}$C [17]. QFMLG was obtained by subsequently annealing the buffer layer samples in the same RF-furnace in a molecular hydrogen atmosphere at about 1 bar and temperature of 800$^{o}$C [15]. The quality, homogeneity, and precise thickness of the buffer layer and QFMLG layers were assessed by angle-resolved photoelectron spectroscopy (ARPES), X-ray photoemission spectroscopy (XPS), atomic force microscopy (AFM), and micro-Raman spectroscopy. ARPES and XPS results for the buffer layer and QFMLG samples are in line with [15] and [17] and confirm the validity of the preparation protocols.

Raman measurements were performed by means of a custom setup. The 488-nm line of an Ar laser at 1.5 mW was focused on the sample by a 0.7-numerical-aperture lens, giving a sub-micron illumination spot. Spatially-resolved Raman spectra were correlated to AFM measurements to assess the distribution of graphene layers on the samples. To this end we exploited the shape of the terraces typically present on the SiC substrate as optical markers. To make these visible, we added an off-optical-axis illumination LED to our system in order to enhance the brightness of the step edges with respect to the flat areas of the substrate. The image of the sample showed well-resolved steps which were used to determine where on the substrate the Raman spectra were actually collected and to execute AFM within the same regions in the following steps. AFM imaging was performed by a Caliber (Veeco) instrument configured in intermittent-contact mode, where topography and phase-images were simultaneously acquired. It was possible to acquire high resolution images of the graphene grown on SiC terraces and to distinguish between the buffer layer and monolayer graphene both by topography and by phase contrast [18].

STM and STS measurements were performed in a variable-temperature ultra high vacuum RHK Technology STM with a base pressure of 5 x 10$^{-11}$ mbar with tungsten tips electrochemically etched in a 2M NaOH solution with homebuilt electronics. In situ, the tips are degassed by placing them near a hot filament overnight. Subsequently, the tips are flashed by applying 600V between tip (positive) and filament (negative) and then quickly increasing the filament current until a 10μA emission current is detected. This removes the oxide from the tips, and then they can be used for imaging. Samples were kept at room temperature, and all measurements were taken in constant-current mode with a tunneling current of 0.3 nA. The WSxM software package was used to analyze STM images [19].

## 3. Results and discussion

Combined Raman and AFM measurements provide valuable spatial information on where to measure the buffer layer and the QFMLG via STM and will be discussed first. Spatially-resolved Raman spectra on the buffer layer regions show the typical SiC bands and no graphene-related bands. However, at selected positions the typical 2D band of monolayer graphene was also detected, characterized by a single Lorentzian peak centered at 2760 cm$^{-1}$ with an average full-width at half maximum (FWHM) of ≈55 cm$^{-1}$, which suggests the presence of inhomogeneous compressive strain [20,21]. Figure 1(a-c) shows the Raman signal (panel (a)), the optical image of the sample (b), and the AFM phase (c) signals obtained at the same position on the sample. The combined Raman and AFM measurements show that monolayer graphene is only present in the proximity of substrate steps. This is consistent with ARPES and XPS data for the same sample (not shown) that indicate a low coverage of monolayer graphene. Away from step edges, the sample is covered by only the buffer layer, which in fact does not produce graphene-related Raman signals. Hydrogen-intercalated samples were also characterized by Raman spectroscopy and AFM. Spatially-resolved Raman data (shown in Fig. 1e) demonstrate the presence of QFMLG throughout the sample (2D peak centered at 2663 cm$^{-1}$, with a FWHM ≈ 30 cm$^{-1}$) [22], except for regions close to the substrate steps (optical image in Fig. 1d) where multilayer graphene, characterized by a broader, blue-shifted 2D band, was observed.

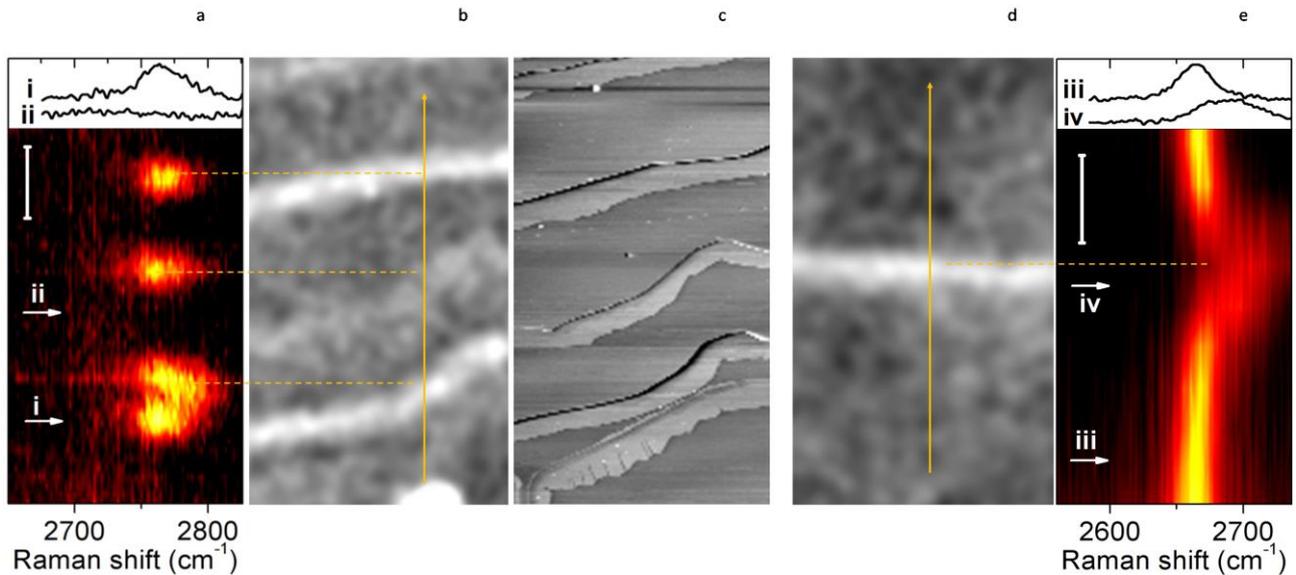

**Fig. 1 - (a) Raman spectra as a function of the position of the excitation-laser spot on the buffer layer sample. The scale bar is 2 μm. As the buffer layer produces no graphene-like Raman signals, the spectra reveal the presence of monolayer graphene inclusions, signaled by the occurrence of the 2D band, only in the proximity of substrate steps. Two spectra collected at the positions marked with "i" and "ii" are shown at the top of panel (a). Step positions were determined by comparing the optical image of the sample (b) and the tapping-mode AFM phase signal (c). Panels (d) and (e) show an optical image and the Raman spectra of one of the QFMLG samples across a step. The scale bar is 1 μm. The 2D band is visible throughout the scan range (yellow arrow in Panel (d)) showing a monolayer character on the terraces and bilayer inclusions in the proximity of the SiC step. Two spectra collected at the positions marked with "iii" and "iv" are shown at the top of the panel (e).**

STM was performed within the central terrace regions (away from step edges) where Raman and AFM analyses confirm the presence of the buffer layer in the sample surface. Figure 2a shows images of the buffer layer taken with a sample tunneling bias $U_T = +1.7$ V where the long-range periodicity of the surface is easily observed in STM [8,14]. The main plot in Fig. 2a shows the surface imaged under tunneling conditions where – already on this large scale – atomic resolution of the buffer layer can be recognized. This situation was attained only after imaging the same surface area for an extended period of time (several hours) and most likely as a result of the achievement of stable tip and tunneling conditions. The tunneling conditions typically remain stable for about 30 minutes (for comparison, it takes about 5 to 10 minutes to measure an STM image). At earlier stages, images such as those shown in the inset of panel (a) and in previous works [4,5,8,14,23-25] were obtained. When the tunneling conditions stabilize and the resolution improves, the images show more details and the quasi-(6x6) superstructure is resolved. This accounts for the variation of the buffer layer images observed in literature [4,5,8]. The micrographs shown in panel (a) are dominated by the (6×6) corrugation which is indicated in the main image by the solid diamond together with the $(6\sqrt{3}\times6\sqrt{3})R30^o$ unit cell as a dashed diamond. The measured unit vector lengths of the two periods are 1.85 nm and 3.2 nm, as expected [8]. The strong and easily imaged corrugation of the surface is a consequence of the covalent bonds that form between approximately 30% of the buffer layer carbon atoms and the silicon atoms of the SiC(0001) surface [10,11], as sketched in the inset to Fig. 2b. The buffer layer is smoothly varying on an atomic scale with a superstructure due to the covalent bonds to the substrate, because the out of plane displacement of single atoms is limited by their covalent bonds to the three neighboring carbon atoms; it is energetically favorable to distribute the strain on several neighboring bonds. In addition to the hexagonal periodicity with a large unit cell typically observed for the buffer layer [8,12,13,14], the STM image in panel (a) clearly resolves an additional periodicity that displays a graphene-like atomic arrangement. The atomic structure of the buffer layer is fully resolved in Fig. 2b, a close up STM image obtained at a sample bias of -0.223 V. The honeycomb structure has a measured lattice constant of 2.5 Å ± 0.1 Å clearly showing a graphene-like topography (2.46 Å). Some atomically resolved images of the buffer layer also show the $(6\sqrt{3}x6\sqrt{3}R30°)$ unit cell. We note that resolving the honeycomb structure was also possible at higher biases but never for values of $|U_T| < 200$ mV. Tunneling conditions became unstable below this sample bias, as expected for the buffer layer [24]. We also point out that atomically-resolved images such as the one reported in panel (b) were consistently acquired when measuring within the terraces in several different areas of the sample. These first STM images of the buffer layer with full atomic resolution clarify the debate concerning the atomic structure of the buffer layer. They demonstrate that the buffer layer is topologically identical to monolayer graphene and thus represents a true periodic carbon honeycomb structure.

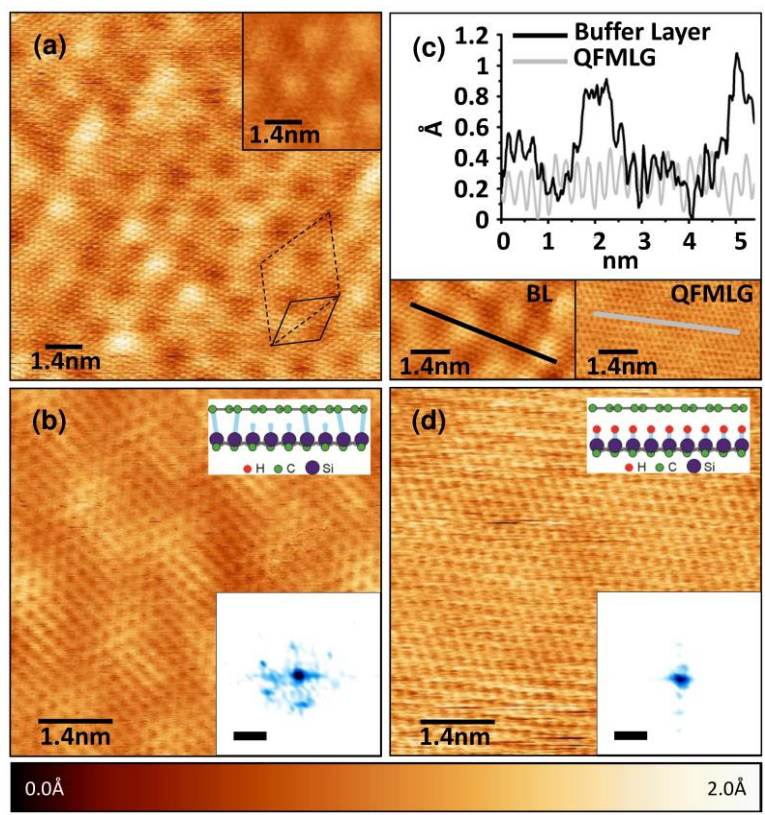

**Fig. 2 - STM images of (a-b) the buffer layer and (d) QFMLG. Panel (a) shows the long-range periodicity imposed on the buffer layer by the substrate. The solid (dashed) diamond designates the (6×6) ((6√3×6√3)R30°) unit cell. Images in panel (a) were taken with a sample bias of +1.7 V. Under optimal tunneling conditions (main image in panel a) as opposed to earlier stages (inset in (a)) the atomic lattice superimposed on the (6×6) periodicity is revealed. Panels (b) and (d) are zoomed-in images of the buffer layer and QFMLG imaged with a sample bias of -0.223 V and +0.103 V, respectively. The upper insets in (b, d) present the**

**structural models of the buffer layer and QFMLG, respectively. The lower insets in panels (b) and (d) are are zoomed in 2D Fast Fourier transforms (2DFFT) of one of the (1x1) spots of the graphene lattice with the quasi-(6x6) satellite spots visible only on the buffer layer. Scale bar 0.58 nm$^{-1}$. Panel (c) shows atomically resolved STM images taken on the buffer layer and QFMLG and the corresponding line profiles along the graphene periodicity. The STM images in Panel (c) have been filtered to remove noise. All measurements were taken in constant current mode with the current set to 0.3 nA.**

Figure 2d shows an STM image of QFMLG obtained at a sample bias of +0.103 V, a bias that yields clear atomically resolved images on monolayer graphene but at which the buffer layer cannot be imaged [4,8,14,24,26]. From these images we extract a lattice constant of 2.4 Å ± 0.1 Å consistent with that of pristine graphene. The (6×6) periodicity present on the buffer layer disappears after intercalation of hydrogen, and the layer appears flat. Also, no obvious atomic defects can be observed in Fig. 2d or in other QFMLG images. This indicates that the process of hydrogen intercalation is rather gentle, and despite the high process temperature additional atomic defects are not noticeably introduced into the graphene layer.

Figure 2c illustrates a roughness analysis of the two surfaces. For this purpose the images were filtered with a Gaussian profile of decay length 3 pixels (1 pixel ~ 0.18 Å) to reduce noise. A line-profile analysis of the buffer layer image is shown by the black line in panel (c) and demonstrates the high corrugation of this interface layer which is a result of the spatially-varying coupling to the SiC substrate mediated by covalent bonds as noted above. The peak-to-peak corrugation value of ~1 Å agrees with what was theoretically calculated by Varchon et al. [12]. The root mean square (RMS) roughness value, calculated from 5 buffer layer images of 25 nm$^2$ each, is 0.178 Å ± 0.020 Å. As shown by the light gray line in panel (c), in the case of the QFMLG sample, the line profile analysis yielded a peak-to-peak corrugation of approximately 0.4 Å. Also in this case the RMS value was calculated from 5 QFMLG images: we obtained a value of 0.125 Å ± 0.005 Å, demonstrating that QFMLG is flatter than the buffer layer. We caution the reader that both the long-range corrugation due to the reconstruction and the atomic corrugation contribute to the peak-to-peak values, so that the real differences in reconstruction-related corrugation are actually more pronounced. For the buffer layer, the corrugation due to the $(6\sqrt{3}\times6\sqrt{3})R30^o$ reconstruction alone amounts to approximately 0.6 Å. On the other hand, for QFMLG the peak-to-peak corrugation is dominated by the graphene lattice (~ 0.3 Å) while the residual long-range variations are around 0.1 Å.

From 2D Fast Fourier transforms (2DFFT) of the STM images additional information on the periodicity of the crystalline structure can be obtained. A single (1x1) spot from the graphene-like lattice is shown in the 2DFFT (Fig. 2b inset) surrounded by 6 satellite spots from the quasi-(6×6) periodicity of the buffer layer. The satellite spots are at a distance of 0.59 nm$^{-1}$ ± 0.07 nm$^{-1}$ which corresponds to 16.9 Å ± 1.5 Å in real space, in close agreement with the expected values (0.54 nm$^{-1}$ and 18.5 Å) [8,26]. At the same time the absence of satellite spots in the QFMLG 2DFFT (Fig. 2d inset) confirms the absence of a quasi-(6×6) periodicity in the real-space images.

To further demonstrate that the atomic resolution images in Fig. 2(a-c) were obtained on the buffer layer and not on minor inclusions of monolayer graphene, STS was performed on the same areas imaged by STM. Figure 3a shows the average of multiple I-V curves acquired in various points on the buffer layer (red line) and on the QFMLG (blue line). The extremely low currents measured for tunneling voltages in the range from -0.5 V to +0.5 V (see red line in Fig. 3a) confirm that Figs. 2(a-c) were taken on buffer layer areas. A low density of states in the vicinity of the Fermi level is a consequence of the strongly-modified electronic structure of the buffer layer due to partial hybridization of its carbon atoms with the SiC substrate [10,11]. Figure 3b shows the differential conductance spectra, *i.e.* the derivative of the I-V curves plotted in panel (a). The buffer layer spectra (red line in Fig. 3b) show virtually no conductance over an energy range of approximately ±0.5 eV with respect to the Fermi level. In stark contrast, QFMLG samples exhibit graphene-like semimetallic differential conductance (blue line in Fig. 3b). The intercalation of hydrogen lifts the electronic coupling of the buffer layer to the substrate and changes its electronic structure to a pristine graphene-like character. The reported dI/dV curves are qualitatively similar to those obtained by Rutter et al. for the buffer layer and monolayer graphene [24]. The dI/dV curves of the QFMLG display a minimum near zero sample bias but the value is finite and does not vanish similarly to the observation by Lauffer et al. [14] for as-grown monolayer graphene on SiC(0001). QFMLG appears to be slightly p-type doped as the minimum of the dI/dV curve *i.e.*, the Dirac point, is shifted to a positive sample bias of about 13 mV (Fig. 3b). Further STS experiments at low temperatures might explore the impact of electron-electron interactions as predicted in [27].

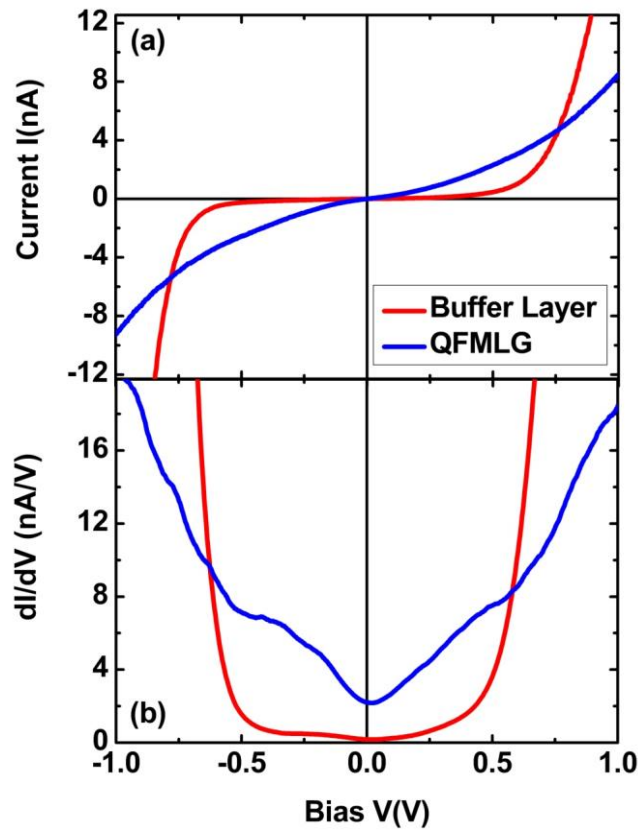

**Fig. 3 - (a) Current vs. voltage (I-V) curves and (b) differential conductance spectra acquired on the buffer layer (red line) and on QFMLG (blue line). The I-V curves in (a) are an average of multiple curves. The spectrum of the buffer layer reveals a low density of states ranging from around -0.5 V to +0.5 V, whereas hydrogen intercalation restores the semimetallic behavior of QFMLG expected for pristine graphene.**

## 4. Summary

We showed by atomically-resolved STM imaging that the buffer layer is topologically identical to a graphene monolayer and thus represents a true periodic carbon honeycomb structure. Furthermore, we compared the atomic structure of the buffer layer to that of QFMLG obtained by hydrogen intercalation. We showed that the (6×6) corrugation existing on as-grown buffer layer samples disappears upon hydrogen intercalation, and that the resulting QFMLG is extremely flat. Also, no obvious atomic defects were observed in the images taken on QFMLG. On the buffer layer, the local density of states as measured via STS is extremely low in an energy range of about ±0.5 eV around the Fermi level, consistently with a partly $sp^3$-hybridized graphenic layer. Only upon hydrogen intercalation does the semimetallic differential conductance spectrum re-emerge. Remarkably, all the results reported in this work were obtained at room temperature. The possibility of observing the atomic structure of the buffer layer and of QFMLG via STM is instrumental for the optimization of growth and intercalation processes. Also, atomic-resolution STM imaging of the buffer layer and QFMLG paves the way for adsorption studies on such layers [28].

## Acknowledgement

We would like to thank A. Bifone of CNI @ NEST (Pisa, Italy) for useful discussions. One of us (V. Pellegrini) acknowledges support from the Italian Ministry of Education, University, and Research (MIUR) through the program "FIRB - Futuro in Ricerca 2010" (project title "PLASMOGRAPH"). This work was supported by the German Research Foundation (DFG) in the framework of the Priority Program 1459 *Graphene*.

## References

[1]   First PN, de Heer WA, Seyller T, Berger C, Stroscio JA, Moon JS. Epitaxial Graphenes on Silicon Carbide. MRS Bulletin 2010; 35: 296-305.

[2] Emtsev KV, Bostwick A, Horn K, Jobst J, Kellogg GL, Ley L, et al. Towards wafer-size graphene layers by atmospheric pressure graphitization of silicon carbide. Nature Mater. 2009; 8: 203-207.

[3] Virojanadara C, Syväjarvi M, Yakimova R, Johansson LI, Zakharov AA, Balasubramanian T. Homogeneous large-area graphene layer growth on 6*H*-SiC(0001). Phys. Rev. B 2008; 78: 245403.

[4] Starke U, Riedl C. Epitaxial graphene on SiC(0001) and SiC(000-1): from surface reconstructions to carbon electronics. J. Phys.: Condens. Matter 2009; 21: 134016.

[5] Chen W, Xu H, Liu L, Gao X, Qi D, Peng G, et al. Atomic structure of the 6H–SiC(0001) nanomesh. Surf. Sci. 2005; 596: 176-186.

[6] Li L, Tsong IST. Atomic structures of 6H-SiC (0001) and (000-1) surfaces. Surf. Sci. 1996; 351: 141-148.

[7] Mårtensson P, Owman F, Johansson LI. Morphology, Atomic and Electronic Structure of 6H-SiC(0001) Surfaces. Phys. Status Solidi B 1997; 202: 501-528.

[8] Riedl C, Starke U, Bernhardt J, Franke M, Heinz K. Structural properties of the graphene-SiC(0001) interface as a key for the preparation of homogeneous large-terrace graphene surfaces. Phys. Rev. B 2007; 76: 245406.

[9] Van Bommel AJ, Crombeen JE, Van Tooren A. LEED and Auger electron observations of the SiC(0001) surface. Surf. Sci. 1975; 48: 463-472.

[10] Emtsev KV, Speck F, Seyller T, Ley L, Riley JD. Interaction, growth, and ordering of epitaxial graphene on SiC{0001} surfaces: A comparative photoelectron spectroscopy study. Phys. Rev. B 2008; 77: 155303.

[11] Riedl C, Coletti C, Starke U. Structural and electronic properties of epitaxial graphene on SiC(0001): a review of growth, characterization, transfer doping and hydrogen intercalation. J. Phys. D: Appl. Phys. 2010; 43: 374009.

[12] Varchon F, Mallet P, Veuillen JY, Magaud L. Ripples in epitaxial graphene on the Si-terminated SiC(0001) surface. Phys. Rev. B 2008; 77: 235412.

[13] Kim S, Ihm J, Choi HJ, Son Y-W. Origin of Anomalous Electronic Structures of Epitaxial Graphene on Silicon Carbide. Phys. Rev. Lett. 2008; 100: 176802.

[14] Lauffer P, Emtsev KV, Graupner R, Seyller T, Ley L, Reshanov SA, et al. Atomic and electronic structure of few-layer graphene on SiC(0001) studied with scanning tunneling microscopy and spectroscopy. Phys. Rev. B 2008; 77: 155426.

[15] Riedl C, Coletti C, Iwasaki T, Zakharov AA, Starke U. Quasi-Free-Standing Epitaxial Graphene on SiC Obtained by Hydrogen Intercalation. Phys. Rev. Lett. 2009; 103: 246804.

[16] Qi Y, Rhim SH, Sun GF, Weinert M, Li L. Epitaxial Graphene on SiC(0001): More than Just Honeycombs. Phys. Rev. Lett. 2010; 105: 085502.

[17] Forti S, Emtsev KV, Coletti C, Zakharov AA, Riedl C, Starke U. Large-area homogeneous quasifree standing epitaxial graphene on SiC(0001): Electronic and structural characterization. Phys. Rev. B 2011; 84: 125449.

[18] Tamayo J, Garcia R. Deformation, Contact Time, and Phase Contrast in Tapping Mode Scanning Force Microscopy. Langmuir 1996; 12: 4430-4435.

[19] Horcas I, Fernandez R, Gomez-Rodriguez JM, Colchero J, Gomez-Herrero J, Baro AM. WSXM: A software for scanning probe microscopy and a tool for nanotechnology. Rev. Sci. Instrum. 2007; 78: 013705.

[20]   Schmidt DA, Ohta T, Beechem TE. Strain and charge carrier coupling in epitaxial graphene. Phys. Rev. B 2011; 84: 235422.

[21]   Röhrl J, Hundhausen M, Emtsev KV, Seyller T, Graupner R, Ley L. Raman spectra of epitaxial graphene on SiC(0001). Appl. Phys. Lett. 2008; 92: 201918.

[22]   Speck F, Jobst J, Fromm F, Ostler M, Waldmann D, Hundhausen M, et al. The quasi-free-standing nature of graphene on H-saturated SiC(0001). Appl. Phys. Lett. 2011; 99: 122106.

[23]   Owman F, Mårtensson P. The SiC(0001)6√3 × 6√3 reconstruction studied with STM and LEED. Surf. Sci. 1996; 369: 126-136.

[24]   Rutter GM, Guisinger NP, Crain JN, Jarvis EAA, Stiles MD, Li T, et al. Imaging the interface of epitaxial graphene with silicon carbide via scanning tunneling microscopy. Phys. Rev. B 2007; 76: 235416.

[25]   Mallet P, Varchon F, Naud C, Magaud L, Berger C, Veuillen JY. Electron states of mono- and bilayer graphene on SiC probed by scanning-tunneling microscopy. Phys. Rev. B 2007; 76: 041403.

[26]   Veuillen JY, Hiebel F, Magaud L, Mallet P, Varchon F. Interface structure of graphene on SiC: an *ab initio* and STM approach. J. Phys. D: Appl. Phys. 2010; 43: 374008.

[27]   Principi A, Polini M, Asgari R, MacDonald AH. The tunneling density-of-states of interacting massless Dirac fermions. Solid State Commun. 2012; 152: 1456-1459.

[28]   Tozzini V, Pellegrini V. Reversible Hydrogen Storage by Controlled Buckling of Graphene Layers. J. Phys. Chem. C 2011; 115: 25523-25528.